\documentclass[times]{tr}

\newif\ifanonymous \anonymousfalse

\usepackage{a4wide}
\usepackage{natbib}
\usepackage{graphicx}
\usepackage[raggedright,tabbotcap]{subfigure}
\usepackage{url}
\usepackage{multicol}
\usepackage{xspace}

\ifanonymous
\newcommand{\studentvillage}{\url{socialsite.xxx}}
\newcommand{\wits}{Anonymous Institution\xspace}
\else
\newcommand{\studentvillage}{\url{studentvillage.co.za}}
\newcommand{\wits}{University of the Witwatersrand, Johannesburg (Wits)}
\fi

\ifanonymous
\author{Several Authors}
\title{An empirical analysis of
  the relationship between web usage and academic performance in
  undergraduate students\\Anonymous Institution}
\else

\title{What clever hominids browse: 
An empirical analysis of
  the relationship between web usage and academic performance in
  undergraduate students}

\author{Scott Hazelhurst}
\author{Yestin Johnson}
\author{Ian Sanders}
\fntext[fn1]{The authors are with the University of the Witwatersrand, Johannesurg. Email addresses: Scott.Hazelurst@wits.ac.za;
  Ian.Sanders@wits.ac.za. Hazelhurst phone and fax:  +27 11 717-6181  fax: +27 865536657 }
\fi

\newcommand{\supp}{the appendix\xspace}

\begin{document}

\begin{abstract}
The use of the internet, and in particular web browsing, offers many
potential advantages for educational institutions as students have
access to a wide range of information previously not available. However,
there are potential negative effects due to factors such as
time-wasting and asocial behaviour.

In this study, we conducted an empirical investigation of the academic
performance and the web-usage pattern of $2153$ undergraduate students.
Data from university proxy logs allows us to examine usage patterns 
and we compared this data to the students' academic performance.

The results show that there is a small but significant (both
statistically and educationally) association between heavier web browsing
and poorer academic results (lower average mark, higher failure
rates). In addition, among good students, the proportion of
students who are relatively light users of the internet is
significantly greater than would be expected by chance.
\end{abstract}

\maketitle

\section*{Citation}
\hrulefill

\noindent

{\large
This is an extended version of a paper that appeared in the 2011
Southern African Computer Lecturer's Association Conference. If you
reference this work please cite:
\begin{itemize}
\item S. Hazelhurst, Y. Johnson, I. Sanders. An empirical analysis of the relationship between web usage and academic performance in undergraduate students. \emph{Proceedings of the Annual Conference of the South African Computer Lecturer's Association}, Ballito, South Africa, July 2011, pp. 29-37.
\end{itemize}
}
\hrulefill

\section{Introduction}

The widespread availability of resources on the internet and their
potential uses in educational settings has driven much debate in their
use for teaching and learning. Students have easier access to a wider
range of material, and can draw links between different information in
new ways. However, the use of the web has been associated with
negative behaviours and outcomes. Now that the internet has been used
in universities for 15 years, and we have a generation of students who
grew up with the internet, we are able to measure the impact and
explore what type of web browsing behaviour is beneficial for
students.
\ifanonymous
\else
\footnote{The title of our paper is drawn from a 1982 paper
  in which \citet{owensmith82} explored the foraging habits of Kudus
  and other ungulates. The choice of title is not just whimsical, as
  we use a technique from this paper to help analyse the internet
  usage of students.}
\fi

The internet has many different uses in a teaching and learning
environment. We focus on the use of the world wide web -- essentially
the use of the \url{http} protocol -- and investigate the association
of academic performance and use of the web in a group of second year
students at \wits. There have
been a number of previous studies on this subject; however, they have
relied on qualitative assessment of internet usage as well as
self-reporting. These methodologies have a number of advantages, but
they can only measure internet usage crudely. In our study, we have
been able to draw on detailed proxy logs of student use of the
internet. In particular, we can be sure that the logs we have for
students who live in the University residences are a complete picture
of the students' internet usage.

The results of the study can be summarised as follows:

\begin{enumerate}
\item There is a small but significant -- in both a statistical and
  educational sense -- association between increased internet usage
  and poorer academic performance.
\item There is a distinct difference in usage patterns between good
  and weak students -- the proportion of light users of the internet
  among good students is much higher than would be expected by chance.
\end{enumerate}

We emphasise the use of the words ``correlation'' and ``association''
rather than ``causality''. First, studies such as we have done can
only show correlation, not causality. Second, even if there is
causality --- high internet usage causes poorer performance --- it is
likely that a deeper causality is more important: For example,
depression is associated with both poorer academic
performance \cite{eisenberg2009} and higher internet usage
\cite{morrison2010}. Moreover, we cannot exclude that merely
suppressing dysfunctional use of the internet would result in
other dysfunctional behaviour. Thus, we see high internet usage as a
symptom rather than the problem.

\subsubsection*{Structure of paper}

We begin by discussing related research in Section
\ref{sec:related}. Section \ref{sec:method} presents the
research methodology. Section \ref{sec:use} gives an overview of
internet usage, which is followed by the key research results in
Section~\ref{sec:results}. In Section \ref{sec:other} we briefly
discuss other investigations that we performed. Finally,
Section~\ref{sec:concl} discusses the results and conclusions.

The analysis of the results are too lengthy for full inclusion in this
paper. We have selected some key results -- the full results can be
found in \supp, as well as in an initial study %
\ifanonymous
[anon ref: student report].
\else
done by \citet{johnson2009}.
\fi 

\section{Related work}

\label{sec:related}

The internet offers incredible access to resources for work, study and
entertainment. Over the last 15 to 20 years much research has 
focussed on the use of the internet by various sectors of
society. These papers have considered a range of subjects
(adolescents, university students, disabled persons) and dealt factors
like access, gender differences, social impacts etc.\ --- see for example
\citep{Hoffmanetal1996,Kaye2000,SchumacherMorahanMartin2001,Singh2001,Gross2004,Erdogan2008,TsaiTsai2010,Valckeetal2010} ---
and various authors have studied \emph{internet addiction} or
\emph{pathological Internet use} in the general population,
e.g., \cite{Young1996,Young1999,Armstrongetal2000,Davis2001,MorahanMartin2005}.

As mentioned above the resources of the Internet give enormous scope
for a richer academic experience for students but also potentially
offer a vast range of distractions which could impact negatively on
the academic performance of both school and university students. A
number of papers have focussed on the effect of Internet use on the
academic performance of school children or adolescents
(e.g. \cite{Hunleyetal2005,Jacksonetal2006,Lubans1999,Witt:07,PierceVaca2008,Mythilyetal2008,GilFlores2009})

\paragraph*{General impact of internet use }

Internet use (or abuse) by university students has been one focus of
research. Some research focuses on general Internet use by students
(e.g. \cite{Odelletal2000} which looks at gender differences in
Internet use and \cite{Korgenetal2001,CottonJelenewicz2006} which
both consider race/ethnicity differences). Some research considers how
and/or why students use the {Web/Internet}. For example,
\citet{Perryetal1998} surveyed 548 students from 3 universities to see
how many students regularly use the Internet, how many hours per week
regular users spend on the Internet and what computers they use. They
also asked respondents their views of their future use of the Internet
in their future careers. \citet{Rumbough2001} investigated
controversial uses of the {Internet} by university students
(e.g. academic cheating, fake emails, pornography,
etc.). \citet{Metzgeretal2003} found that college students report that
they rely very heavily on the {Web} for general and academic
information; that their use includes research (getting information)
for school work, banking and stock market information, email, checking
sports scores and downloading music; and that they believe that this
use will increase over time).

\citet{Gord:07} investigated internet use and well-being among college
students, with focus on frequency of use. This study aimed to
determine what students use the internet for and how each of these
affect their performance in college. A survey was performed on a
representative sample of undergraduate students. This study identifies
the top five types of internet use reported by students in the sample.
The five types identified were: emailing friends, getting help with school work,
talking with friends, emailing family, and instant messaging. These uses
did not differ significantly between gender. Frequency of internet use was
not found to be correlated with any of the well-being measures.
It was found that the amount of time spent online was significantly 
associated with social anxiety, however this association became marginal
after the types of use were entered into the model. The findings in this
study suggest that the specific type of internet use relates to depression,
social anxiety and family cohesion much more so than does frequency of use.
It was also found that the internet has become an important aspect of
college students' lives. It was revealed that students mainly used the 
internet to email family and friends, IM, talk with friends, and get help
with school work. This shows that students were drawn to the Internet
primarily as a means of communication with friends and family. These
results are similar to those found in previous studies. It was also
found that men use the internet more for leisure, while women use the
internet more often for communication. However, not much else was found
in the way of gender differences. This shows that gender differences
in patterns of internet use may be relatively small. The relationship
between internet use and well-being is complex, what matters to a 
student' well-being is not necessarily how long they spend online,
but what they do online.

\citet{Fortsonetal2007} reported on internet use, abuse and dependence among
students at a regional U.S. university. Once again a survey was used to
gather information about the students in the sample. It was found that
the majority of students use the internet daily, and that half of the
sample met the defined criteria for internet abuse. There were no gender
differences in terms of daily access to the internet, however males
and females did seem to use the internet for different
reasons. Finally, depression was found to be positively correlated with more
frequent internet use \cite{Fortsonetal2007}.

A concern which is prevalent in the literature is whether
``excessive'' {Internet} use (also termed {Internet} addiction or
pathological {Internet} use) could have a negative effect on the
academic experience of university students. Some researchers report
negative effects of {Internet Addiction} such as increasing time spent
on line, disturbances of sleep patterns, isolation, etc. which could
have an effect on academic performance but do not directly address the
issue of academic performance (see for example
\citet{Kandell1998,ChouHsiao2000,MorahanMartinSchumacher2000,Chou2001,Rotsztein2003,Fortsonetal2007,OdaciKalkan2010}). 

Another area of research concentrates on the adoption of the
{Internet} by institutions and focuses on appropriate adoption/use
strategies and the effect of the adoption of the technology on the
students' university experience (see for example
\citet{Jones2002,MatthewsSchrum2003,Hongetal2003,CheungHuang2005,SalaamAdegbore2010}). Some
of this research specifically considers the effect of the adoption of
the Internet on students' academic performance. For example
\citet{Osunadeetal2009} shows that there is a significant difference
in academic performance between students at institutions that have
Internet infrastructure and access on their campus and those that do
not; and \citet{Tella2007} who studied the Internet usage of
undergraduate students in {Botswana} shows that most of their
respondents reported using the Internet for the purpose of obtaining
course related information and that the Internet contributes
significantly to their academic performance.

\paragraph*{Academic performance} \citet{Kube:01} present the early
findings of how internet use affects collegiate academic
performance. This study focuses on students' dependence on the
internet and attempts to quantify to what extent students are addicted
to the internet. It was found that a significant percentage of
students whose academic performance was bad indicated that the
internet kept them up late at night, thereby making them tired for
lectures the following day \cite{Kube:01}. Strong evidence was found
to suggest that students' excessive use of the internet is associated
with academic problems, however it was unclear if these students would
have had similar problems even without the internet being so readily
available. These findings demonstrate the need for more research into
this field, and specifically the need for hard data to be analysed,
and compared to the self-reported data available from other surveys.

\citet{Witt:07} explored whether a students' computer use at home is
related to their mathematical performance at school revealed some
important results. The research aimed to determine if a student using
the internet at home would have a different mathematical performance
in school than a student with no internet access. An important aspect
of this research is that it compares the effects of home internet
usage to other factors that have been identified as being important in
prior studies. These factors include immigration background, leisure
activities, cognitive abilities, how often they read books and
newspapers, how often they watch television and the news, and how
often they watch horror, action, or pornographic films. It was found
that overall a student's computer-related behaviour at home only plays
a marginal role in predicting their academic performance. It was found
that when compared to a student's cognitive abilities, their
immigration background, and leisure activities other than computer
use, that a student's access to a home computer did not contribute
towards explaining differences in their mathematical performance.  The
frequency of computer use was also found to not contribute towards
mathematical performance. However, other studies have found
contradictory results, but measures taken in this research attempt to
ensure the accuracy of these results. Finally, the only major computer
related factor contributing towards students' mathematical performance
was found to be if the student was particularly interested in
computers, and had acquired their computer skills themselves.

Other work more directly addresses the relationship between ``high''
Internet use and academic performance. \citet{Scherer1997} reports
that 13\% of the respondents in her study reported ``excessive"
{Internet} use that interfered with personal functioning.  The study
by \citet{Anderson2001} presents similar results for a small group of
students (106 from a total sample of 1300 students from eight academic
institutions) who used the {Internet} ``excessively'' -- these
students were significantly more likely to indicate that their
{Internet} use negatively affected their academic performance, meeting
new people and their sleep patterns.  \citet{SuhailBargees2006}
surveyed 200 undergraduate students in Pakistan and found that
``excessive'' {Internet} use can lead to many problems -- educational,
physical, psychological and interpersonal.  \citet{ChenPeng2008}
surveyed a large sample of students in Taiwan and found that students
who reported ``heavy'' {Internet} use were more likely to have worse
academic grades, have worse relationships with the administrative
staff, lower learning satisfaction and to be depressed, physically
ill, lonely and introverted.

\citet{Frangosetal2010} studied a sample of 1876 Greek university
students in order to establish the degree of Internet addiction in
these students. They applied a Greek version of Young's Internet
Addiction survey and added items on demographic factors and questions
about academic performance. They found Internet Addicted students were
more likely to report poor academic performance and that Internet
Addiction was predicted by increased hours of daily Internet use;
increased hours visiting chat rooms, sex pages and blogs; being male;
being divorced; having poor grades; and accessing the {Internet}
outside of the home. \citet{EnglanderTerregrossa2010} found a negative
and statistically significant correlation between time spent on line
and the grade performance of 128 students in an introductory
micro-economics course.

\paragraph*{Social networking} A survey performed by \citet{PierceVaca2008}
to determine the differences in performance between teen users and
non-users of social networking sites such as MySpace, as well as of
other communication technologies, reveals some interesting
results. The study made use of a survey which aimed to determine how
many students report having a profile on a social network site, how
many have cellphones, how many use text messaging, and in turn if the
student uses each of these while doing homework, while in class, or
during tests and exams. It was found that teen users of MySpace
reported significantly lower grades than those who did not use the
service. The same is true for teens with an instant messaging account,
those with a cellphone, and those with text-messaging.  A significant
difference was found between students who had a MySpace account and
those who did not -- those who had a MySpace account reported
significantly lower grades than those who did not have a MySpace
account.  A significant difference was also found between those who
reported having an IM account, and those that did not. Finally, a
significant difference was also found between those having a cell
phone, and those without one.  Those with cell phones reported
significantly lower grades than those without one. Those who did have
text-messaging on their cell phones also reported significantly lower
marks than those who did not have text-messaging. It was also found
that those who said they kept their MySpace open while doing homework
reported significantly lower grades than those who did not keep their
MySpace open while doing homework.  The same was true for those who
kept their IM account open while doing homework and those that did
not. Those who text-messaged or who talked on their phones while doing
homework also reported significantly lower grades than those that did
not. It was also found that those who put off doing homework to spend
time on MySpace reported significantly lower grades than those that
did not put off their homework.  While it is not possible for the
results of this study to reveal any causal link between grades and
technology use, they do suggest that certain technologies can be very
distracting to teens. This in turn can be linked to lower grades. It
was also found that many students reported text messaging during
class-time. This suggests that students are not paying as careful
attention in class as they could be. Some students even reported text
messaging during exams, this suggests that teens are using cell phones
as a source of cheating which is highly disturbing.

\paragraph*{Sources of data}
A common theme of the research discussed above is that the data about
Internet usage was collected by means of questionnaires or surveys
administered to the subjects --- i.e., the data used was
self-reported. In the work of \citet{Jacksonetal2006} {Internet} usage
(time online, number of sessions, domains visited and emails sent) was
automatically collected for a period of 16 months. This data was then
related to students' grades. This research focussed on a fairly small
group of low income high school students but the approach is similar
to that which we adopted in our study. There has been no work reported
that looks at hard data reflecting university students' actual online
behaviour and relating that to academic performance.

\section{Methodology}
\label{sec:method}

This section discusses the data and how it was analysed.
\wits is a research university
based \emph{(suppressed)}. There are approximately 18000
students in a first bachelor's degree and about 6000 masters and
doctoral students. There are five faculties: Science; Engineering \&
the Built Environment; Health Sciences; Commerce, Law \& Management;
and Humanities. The student population is very diverse, reflective of
the region's population (gender, race, and class). In particular some
students had little or no exposure to computers and the internet
before they came to University, while others have had computers since
they were small children, with a good internet connection at home and
in their high schools.

\paragraph*{Sample of students} Second-year students were chosen as
the focus of the study to reduce the heterogeneity of the
sample. There were two reasons for doing this: (1) all students who
pass the first year will have shown their ability to succeed at
University and will have had some computing and internet experience,
and (2) students at a more senior level may have significant
discipline-specific internet usage requirements.

We were particularly interested in those students who were residing in
university residences. Most importantly, it is a reasonable
assumption to make that for these students there was no significant
internet usage that was not captured by the proxies -- the relative
cost/performance of the university internet service for students (free,
reasonable bandwidth) versus internet caf\'es and 3G services makes it
unlikely students could use these services, and they would not have
had significant access to DSL services. Another factor is that
students in the \emph{undergraduate} residences are much more
homogeneous with respect to race and class than the general population.

\paragraph*{Academic results} We selected a set of 11 second year
courses across the University and used for our study all the students
registered for these courses. This gives us a range of students in
different disciplines and a sample in each discipline.  We obtained
\emph{all} the marks for these students (not just the marks of the 11
courses) as well as the marks of other students to enable us to
compute the average marks of any courses taken by the student
sample. For each student we also recorded whether they were in a
University residence. We excluded from the study all students who had
obviously dropped out (marks of 0/failed absent).

\paragraph*{Internet usage} We obtained the Squid proxy logs for all the
students in the study for the second half of the academic year. In the
2007 academic year, the University had a strict policy which meant
that all web browsing had to be done through an authenticating
proxy. The log files were about $18.6$ GB in size, with just over
$105$ million entries (URL requests).

\paragraph*{Treatment of data} 
The data was anonymised in such a way that we could link academic
performance and internet usage. It was then imported into a SQLite
database -- most of the analysis was done using SQL; however Python
scripts were used for investigating the number of sessions. R was used for
statistical analysis.

\paragraph*{Categories of student} In the remainder of the paper we
use the categories of the students based on their usage patterns as
given below:
\begin{itemize}
\item Very heavy users: those in the top 10\% of users by usage.
\item Heavy users: those in percentiles 60-90\% of users by usage.
\item Light users: those in percentiles 10-40\% of users by usage.
\item Very light users: those in the bottom 10\% of users by usage
\end{itemize}

\noindent
We study all four categories, though we focus on the heavy/light users rather
than the very heavy/very light users since we are more interested in
general effect rather than extremes.

\paragraph*{Measurement of academic performance} Although we have hard
data on student performance in courses, there is no completely
accurate way to measure overall performance for a student since
workload varies and different courses have markedly different
averages. Ideally we would have one measure of student
performance. However, there is no university determined index (such as
a GPA).

For each student, we had all the courses done in the 2007 academic
year. From this we computed the \emph{weighed average} of the
courses. Assuming a student $i$ did $n_i$ courses, the weighted average is
$\sum_{j=1}^{n_i}p_ic_{i,j}/n_i$, where $p_i$ is the weighting of the course and
$c_{i,j}$ is the mark the student $i$ obtained for course $j$.

From an initial analysis, we could see that internet usage varied
across the different subjects as did the average mark for courses. For
this reason, we also defined a \emph{performance index} (PI) by
relativising the mark the student obtained by the course
average\footnote{Across all students who did the course, not just
  students in the study. Thus the average performance index of all
  students in the study is not exactly 1.}. This then gives a way of
comparing students' relative performance.  Formally the performance
index is $\sum_{j=1}^{n_i}p_i(c_{i,j}-a_j)/\sum_{j=1}^{n_i}p_i$ where $a_j$ is the
average mark of all students in course $j$. 

The weighted average is the more obvious choice, but the performance
index allows us to smooth out differences in standards\footnote{The
  performance index (PI) gives us the difference in percentage points
  between a student and the hypothetical average student, who would
  have a PI of 0. The average PI of the students in the study is just
  less than 0, since the relativisation is done with respect to course
  averages, which included marks of students not covered in the
  study.} (either of the students or marking). One problem with the
performance index is that a student who does a course with relatively
few students can have their PI unrealistically skewed up or down.

The average mark and performance index are both used in the analysis
below because they both capture valid facets of academic
performance. We also look at the proportion of courses passed.

\paragraph*{Measuring internet usage} From the proxy logs, we must
produce a usage index for each student. There are three ways this can
be done. (1) Compute the total number of bytes downloaded; (2) Compute
the number of URLs fetched -- the number of hits -- and (3) Compute
the internet session time. Each of these have their advantages and
disadvantages.

The student's bandwidth utilisation may be misleading since it is
conceivable that a student could write a simple script to download
lots of music, which means they spend little time on the internet. The
number of URLs visited by the student is misleading since most URL
requests are as a result of indirect requests (one web page requesting
others) and this measure may also be skewed by autorefresh.  Measuring
the number and length of sessions is difficult\footnote{We applied the
  techniques used to measure animal behaviour in
  \cite{owensmith82,sibly90}, and consulted the literature
  (e.g. \cite{murray2006}).} for a variety of reasons.


\paragraph*{Students with no proxy records} In this study, we focus on
students for whom we have both academic results and internet proxy
results. However, approximately 28\% of the students (610) had no
proxy records at all. None of these students were in the university
residence; hence, the most likely explanation is that these were
students who had internet connectivity at home and no requirements for
access at the University. The academic performance of these students
was significantly better than the other students -- an average 57.3\%,
and a performance index of $+$1.90. Since we find that lighter
internet use is associated with better performance, excluding this
\emph{noproxy} group of students from the statistical study
strengthens our conclusions.

\section{Overview of internet use}

\label{sec:use}

This section gives an overview of how students used the internet.
\begin{itemize}
\item Section~\ref{sec:use:download} describes usage by number of bytes
  downloaded;
\item Section~\ref{sec:use:hits} describes usage by number of URL
  requests (hits);
\item Section~\ref{sec:use:url} shows which sites were most popular;
\item Section~\ref{sec:use:time} characterises internet use by time of
  day; and 
\item Section~\ref{sec:use:sessions} explores using number of sessions
  as a measurement.
\end{itemize}

\subsection{Internet usage measured by download}

\label{sec:use:download}

Table \ref{table:overview} gives overall statistics of all the students
chosen for the study. As should be expected students in residence use
considerably more university internet resources than students not in
residence. This should not be interpreted as saying that overall they use
the internet more -- it is just a statement about university resource usage.

\begin{table}[ht]\centering
\begin{tabular}{|l r r|}\hline
Category & All students & Res students \\\hline
Number of students selected for study & 2153 & 557\\
Number of selected with results       & 2147 & 533 \\
Number of selected with proxy data    & 1546 & 533 \\
Number with results and proxy data     &    1543 & 533 \\
Average proxy usage  & 494 MB      & 773 MB \\
Maximum proxy usage  & 64 308 MB    & 64 308 MB  \\
Total download       & 762 115 MB   & 412 164 MB\\
Median proxy usage   & 122 MB      & 217 MB\\
Average number of hits (000) & 66.7 & 104\\
Maximum number of hits (000) & 3185  & 3185\\\hline
\end{tabular}
\caption{Overview of students and data usage.}
\label{table:overview}
\end{table}

Table \ref{table:usagebytes:histo} gives a profile of how much downloading
students do (as measured by number of bytes downloaded, including only
those students who used the internet at least once.  In summary, about
two thirds used fewer than 250 MB, three quarters used fewer than 360
MB and just under 1\% used more than 1 GB.  The top 10\% of users use
66\% of total download; the top 20\% use 81\% of total download. For
residence students, the usage figures are not as skewed but still the
top 20\% of users use 66\% of download.

\begin{table}[tb!]\centering
\hfill
\subtable[Breakdown of usage -- total download. Percentages are rounded
  to closest integer. The \emph{Usage} column gives a range (in
  MB). \emph{All students} shows the number and proportion of all
  students whose usage was in this range. \emph{Res students} shows
  the same for students in the residence. For example, 9\% of all
  students and 13\% of residence students used between 200MB and 300MB.]{
\begin{tabular}{|r r r|}\hline
 All students  &  Res students   &  Usage (range in MB) \\\hline
 716 (46\%)      & 144 (27\%)   & [0,100) \\
 249 (16\%)      & 102 (19\%)   & [100,200) \\
 135 \ (9\%)     &  68 (13\%) & [200,300) \\
  83 \ (5\%)     &  42 \ (8\%) & [300,400) \\
  60 \ (4\%)     &  27 \ (5\%) & [400,500)\\
  43 \ (3\%)     &  24 \ (5\%) & [400,500)\\
  39 \ (3\%)     &  20 \ (4\%) & [600,700)\\
  19 \ (1\%)     &   9 \ (2\%) & [700,800)\\
  24 \ (2\%)     &  14 \ (3\%)  & [800,900)\\
  21 \ (1\%)     &  11 \ (2\%)  & [900,1000)\\
  88 \ (6\%)     &  40 \ (8\%)  & [1000,2000)\\
  31 \ (2\%)     &  13 \ (2\%)  & [2000,3000)\\
  11 \ (1\%)     &   5 \ (1\%)  & [3000,4000)\\
   6 \ (0\%)     &   3 \ (1\%)  & [4000,5000)\\
  10 \ (1\%)     &   6 \ (1\%)  & [5000,10000)\\
   8 \ (1\%)     &   5 \ (1\%)  & [10000,65000)\\ \hline  
\end{tabular}
\label{table:usagebytes:histo}}
\hfill
\subtable[Cut-off for percentiles. Entry for Percentile $x$ shows that
  students in the top $x\%$ of internet users in the study used at least this
  amount of data in MB.]{\begin{tabular}{|r| r r|}\hline
Percentile & All students & Res students \\\hline
10   & 1001 & 1338 \\
20   &  482 &  701 \\
30   &  285 &  475 \\
40   &  175 &  312 \\
50   &  122 &  221 \\
60   &  72  &  160 \\
70   &  39  &  120 \\
80   &  16  &   74 \\
90   &   5  &   33 \\ \hline
\end{tabular}
\label{table:percentiles}
}\hfill
\caption{Overview of internet usage by bytes}
\label{table:usagebytes}
\end{table}

Table \ref{table:percentiles} shows the usage patterns of light and
heavy users. For each percentile it shows the number of MB downloaded
by those users. For example, the top 10\% of all students (which we
have characterised as very heavy users) downloaded at least 1001 MB
and the top 10\% of residence students downloaded at least 1338
MB. Similarly, among all students, heavy users downloaded between 175
MB and 1001 MB, light users downloaded between 5 MB and 72 MB, and
very light users downloaded less than 5 MB over the period.

\subsection{Internet usage measured by number of hits}

\label{sec:use:hits}

This sub-section gives a similar overview of internet usage measuring
the number of hits.  Tables \ref{table:numhits} and
\ref{table:hitpercentiles} show the internet usage by number of hits.

\begin{table}[tb!]\centering
\hfill
\subtable[Internet usage measured by number of  hits. This table shows
  what number of users made what number of hits. The range is shown in
  thousands. For example, 217 students made between 10000 and 19999
  hits in the period.]{
\hfill\begin{tabular}{|l|r r|} \hline
Range  & All students& Residence\\\hline
\ [0,10) & 573 & 105\\
\ [10,20) & 217 & 69\\
\ [20,30) & 128 & 57\\
\ [30,40) & 88 & 37\\
\ [40,50) & 85 & 36\\
\ [50,60) & 53 & 24\\
\ [60,70) & 46 & 23\\
\ [70,80) & 36 & 18\\
\ [80,90) & 31 & 10\\
\ [90,99) & 33 & 17\\
\ [100,199) & 144 & 81\\
\ [200,300) & 50 & 21\\
\ [300,400) & 25 & 15\\
\ [400,500) & 11 & 2\\
\ [500,999) & 19 & 12\\
\ [1000,1999) & 4 & 2\\
\ [2000,2999) & 4 & 3\\
\ [3000,3999) & 1 & 1\\\hline
\end{tabular}
\label{table:numhits}
}\hfill
\subtable[Cut-off for percentiles. Entry for Percentile $x$ shows that
  students in the top $x\%$ of internet users made this number of hits.]{\begin{tabular}{|r| r r|}\hline
Percentile & All students & Residence students \\\hline
10   &  156 &  206 \\
20   &   82 &  125 \\
30   &   48 &   82 \\
40   &   31 &   58 \\
50   &   19 &   40 \\
60   &   12 &   26 \\
70   &    6 &   18 \\
80   &    3 &   10 \\
90   &  0.66 &   5 \\ \hline
\end{tabular}
\label{table:hitpercentiles}}\hfill
\caption{Internet usage by number of hits}
\end{table}

\begin{table}[ht!]\centering
\end{table}

\subsection{Analysis of usage by URL}

\emph{Note to reviewers: the details of some URLs is suppressed for
  double-blind reviewing.}
\label{sec:use:url}

\begin{table}[ht!]\centering
\footnotesize
\begin{multicols}{2}

\begin{tabular}{|l@{}l@{}r@{\ }r@{\ }r|}\hline
 & URL & \#hits & Perc. & Cum \\\hline
1 & studentvillage.co.za & 15258 & 14.51 & 14.5\\
2 & facebook.com & 14750 & 14.03 & 28.5\\
3 & google.com & 3011 & 2.86 & 31.4\\
4 & yimg.com & 2862 & 2.72 & 34.1\\
5 & yahoo.com & 2730 & 2.60 & 36.7\\
6 & mtn.co.za & 2381 & 2.26 & 39.0\\
7 & mig33.com & 2175 & 2.07 & 41.1\\
8 & vodacom4me.co.za & 1976 & 1.88 & 42.9\\
9 & hi5.com & 1222 & 1.16 & 44.1\\
10 & google.co.za & 840 & 0.80 & 44.9\\
11 & wm.co.za & 809 & 0.77 & 45.7\\
12 & google-analytics.com & 706 & 0.67 & 46.3\\
13 & slide.com & 624 & 0.59 & 46.9\\
14 & careerjunction.co.za & 620 & 0.59 & 47.5\\
15 & webmail.co.za & 618 & 0.59 & 48.1\\
16 & live.com & 592 & 0.56 & 48.7\\
17 & wits.ac.za & 565 & 0.54 & 49.2\\
18 & standardbank.co.za & 529 & 0.50 & 49.7\\
19 & msn.com & 528 & 0.50 & 50.2\\
20 & person.com & 511 & 0.49 & 50.7\\
21 & thunda.com & 504 & 0.48 & 51.2\\
22 & news24.com & 491 & 0.47 & 51.7\\
23 & chat27.co.za & 451 & 0.43 & 52.1\\
24 & akamai.net & 429 & 0.41 & 52.5\\
25 & com.com & 382 & 0.36 & 52.9\\
26 & iol.co.za & 325 & 0.31 & 53.2\\
27 & bboybunker.com & 312 & 0.30 & 53.5\\
28 & wikimedia.org & 292 & 0.28 & 53.7\\
29 & myspace.com & 271 & 0.26 & 54.0\\
30 & vodacom.co.za & 261 & 0.25 & 54.2\\
31 & rockyou.com & 258 & 0.25 & 54.5\\
32 & mnet.co.za & 254 & 0.24 & 54.7\\
33 & bbc.co.uk & 240 & 0.23 & 55.0\\
34 & myspacecdn.com & 239 & 0.23 & 55.2\\
35 & go.com & 239 & 0.23 & 55.4\\
36 & kaizerchiefs.com & 218 & 0.21 & 55.6\\
37 & images-amazon.com & 216 & 0.21 & 55.8\\
38 & premierleague.com & 206 & 0.20 & 56.0\\
39 & hotmail.com & 206 & 0.20 & 56.2\\
40 & standardbank.co.za & 200 & 0.19 & 56.4\\
41 & gumtree.co.za & 198 & 0.19 & 56.6\\
42 & adbux.org & 196 & 0.19 & 56.8\\
43 & supersport.co.za & 190 & 0.18 & 57.0\\
44 & adinterax.com & 186 & 0.18 & 57.1\\
45 & uct.ac.za & 185 & 0.18 & 57.3\\
46 & cupidbay.com & 183 & 0.17 & 57.5\\
47 & oprah.com & 174 & 0.17 & 57.7\\
48 & hidemylocation.com & 173 & 0.16 & 57.8\\
49 & userplane.com & 172 & 0.16 & 58.0\\
50\phantom{0} & trendmicro.com & 172 & 0.16 & 58.2\\\hline
\end{tabular}

\begin{tabular}{|l@{}l@{}r@{\ }r@{\ }r|}\hline
   & URL & \#hits & Perc. & Cum \\\hline
51 & highveld.co.za & 171 & 0.16 & 58.3\\
52 & pcgames.com.cn & 171 & 0.16 & 58.5\\
53 & quantserve.com & 171 & 0.16 & 58.6\\
54 & jobmail.co.za & 169 & 0.16 & 58.8\\
55 & stickcricket.com & 166 & 0.16 & 59.0\\
56 & 65.111.173.43 & 164 & 0.16 & 59.1\\
57 & wayn.com & 164 & 0.16 & 59.3\\
58 & mnetafrica.com & 164 & 0.16 & 59.4\\
59 & wikipedia.org & 160 & 0.15 & 59.6\\
60 & kickoff.com & 159 & 0.15 & 59.7\\
61 & flixster.com & 148 & 0.14 & 59.9\\
62 & skysports.com & 143 & 0.14 & 60.0\\
63 & absa.co.za & 142 & 0.14 & 60.1\\
64 & 5fm.co.za & 139 & 0.13 & 60.3\\
65 & meetmarket.co.za & 138 & 0.13 & 60.4\\
66 & wwe.com & 137 & 0.13 & 60.5\\
67 & thevoicebw.com & 137 & 0.13 & 60.7\\
68 & atdmt.com & 131 & 0.12 & 60.8\\
69 & findastudent.co.za & 130 & 0.12 & 60.9\\
70 & mg.co.za & 126 & 0.12 & 61.0\\
71 & blogger.com & 126 & 0.12 & 61.2\\
72 & manutd.com & 125 & 0.12 & 61.3\\
73 & adobe.com & 121 & 0.12 & 61.4\\
74 & skurfit.com & 121 & 0.11 & 61.5\\
75 & timeinc.net & 117 & 0.11 & 61.6\\
76 & themafiaboss.com & 114 & 0.11 & 61.7\\
77 & komp3.net & 114 & 0.11 & 61.8\\
78 & tagged.com & 114 & 0.11 & 61.9\\
79 & ebuddy.com & 113 & 0.11 & 62.0\\
80 & 74.86.142 & 113 & 0.11 & 62.2\\
81 & 2o7.net & 108 & 0.10 & 62.3\\
82 & wadja.com & 107 & 0.10 & 62.4\\
83 & cnn.net & 103 & 0.10 & 62.5\\
84 & avoidr.com & 103 & 0.10 & 62.6\\
85 & soundpedia.com & 98 & 0.09 & 62.6\\
86 & uefa.com & 96 & 0.09 & 62.7\\
87 & mweb.co.za & 93 & 0.09 & 62.8\\
88 & vowfm.com & 92 & 0.09 & 62.9\\
89 & yfm.co.za & 91 & 0.09 & 63.0\\
90 & qq.com & 90 & 0.09 & 63.1\\
91 & sowetan.co.za & 89 & 0.09 & 63.2\\
92 & answers.com & 88 & 0.08 & 63.3\\
93 & edumela.com & 88 & 0.08 & 63.3\\
94 & eskom.co.za & 87 & 0.08 & 63.4\\
95 & arsenal-mania.com & 86 & 0.08 & 63.5\\
96 & symantec.com & 86 & 0.08 & 63.6\\
97 & mofunzone.com & 86 & 0.08 & 63.7\\
98 & style.com & 85 & 0.08 & 63.7\\
99 & bizcommunity.com & 85 & 0.08 & 63.8\\
100 & photobucket.com & 85 & 0.08 & 63.9\\\hline
\end{tabular}

\end{multicols}
\caption{Table of top 100 organisational URLs by number of
  hits. \#hits is the
  number of hits  in thousands. \emph{Perc} gives the
  percentage of all hits go to this organisation. \emph{Cum} gives the
cumulative percentage.}
\label{table:urlbyhit}
\end{table}

The analysis of internet usage by URL downloaded is very instructive.
Table \ref{table:urlbyhit} shows internet usage by URL, by number of
hits. 140087 distinct organisational URLs\footnote{To produce this
  table, we automatically stripped URLs so that we produced only
  organisational URLs -- so only \url{.edu} or \url{nih.org} -- no
  sub-domains. We lose some information this way, but otherwise there
  are too many variations to make sense of the data. There is still
  some duplication, e.g. all \url{google} sites, and some just given
  by IP address.} were detected. The top 100 popular sites
collectively account for 63\% of all hits.

\begin{table}[ht!]\centering
\footnotesize
\begin{multicols}{2}
\begin{tabular}{|p{4mm}@{}l@{} r@{\ \ } r@{\ }r|}\hline
   & URL & MB & Perc & Cum \\\hline
1 & facebook.com & 57838 & 7.46 & 7.5\\
2 & yahoo.com & 38999 & 5.03 & 12.5\\
3 & yimg.com & 27966 & 3.61 & 16.1\\
4 & myspace.com & 23005 & 2.97 & 19.1\\
5 & google.com & 17511 & 2.26 & 21.3\\
6 & studentvillage.co.za & 11561 & 1.49 & 22.8\\
7 & narutochaos.com & 10109 & 1.30 & 24.1\\
8 & evilshare.com & 9047 & 1.17 & 25.3\\
9 & hi5.com & 8915 & 1.15 & 26.4\\
10 & symantec.com & 8094 & 1.04 & 27.5\\
11 & wizputer.net & 7772 & 1.00 & 28.5\\
12 & vodacom4me.co.za & 7714 & 1.00 & 29.5\\
13 & badongo.com & 6940 & 0.90 & 30.4\\
14 & wits.ac.za & 6857 & 0.88 & 31.3\\
15 & akamai.net & 5979 & 0.77 & 32.0\\
16 & google.co.za & 5963 & 0.77 & 32.8\\
17 & download.com & 5167 & 0.67 & 33.5\\
18 & nvidia.com & 5104 & 0.66 & 34.1\\
19 & 38.118.213 & 4978 & 0.64 & 34.8\\
20 & 65.111.173 & 4861 & 0.63 & 35.4\\
21 & edgesuite.net & 4815 & 0.62 & 36.0\\
22 & live.com & 4517 & 0.58 & 36.6\\
23 & adobe.com & 4396 & 0.57 & 37.2\\
24 & macromedia.com & 4344 & 0.56 & 37.7\\
25 & ouou.com & 4309 & 0.56 & 38.3\\
26 & go.com & 4295 & 0.55 & 38.9\\
27 & quicksharing.com & 4015 & 0.52 & 39.4\\
28 & mtn.co.za & 3589 & 0.46 & 39.8\\
29 & free.fr & 3473 & 0.45 & 40.3\\
30 & slide.com & 3432 & 0.44 & 40.7\\
31 & pagerealm.com & 3391 & 0.44 & 41.2\\
32 & photobucket.com & 3346 & 0.43 & 41.6\\
33 & msn.com & 3319 & 0.43 & 42.0\\
34 & webmail.co.za & 3300 & 0.43 & 42.4\\
35 & multiply.com & 3037 & 0.39 & 42.8\\
36 & bboybunker.com & 2637 & 0.34 & 43.2\\
37 & vestigialconscience.com & 2598 & 0.34 & 43.5\\
38 & adinterax.com & 2405 & 0.31 & 43.8\\
39 & wikipedia.org & 2240 & 0.29 & 44.1\\
40 & 68.142.200 & 2197 & 0.28 & 44.4\\
41 & 85.131.244 & 2104 & 0.27 & 44.7\\
42 & englers.org & 1984 & 0.26 & 44.9\\
43 & 202.99.174 & 1960 & 0.25 & 45.2\\
44 & careerjunction.co.za & 1878 & 0.24 & 45.4\\
45 & 216.252.110 & 1828 & 0.24 & 45.7\\
46 & ea.com & 1807 & 0.23 & 45.9\\
47 & caltech.edu & 1773 & 0.23 & 46.1\\
48 & ibo.dfx.at & 1748 & 0.23 & 46.3\\
49 & a2zuploads.com & 1683 & 0.22 & 46.6\\
50 & metacafe.com & 1630 & 0.21 & 46.8\\\hline
\end{tabular}

\columnbreak

\begin{tabular}{|p{5mm}@{}l@{}r@{\ \ } r@{\ }r|}\hline
   & URL & MB & Perc & Cum \\\hline
51 & myspacecdn.com & 1630 & 0.21 & 47.0\\
52 & 85.17.172 & 1608 & 0.21 & 47.2\\
53 & onsmashexclus\ldots.to & 1600 & 0.21 & 47.4\\
54 & 38.119.88 & 1595 & 0.21 & 47.6\\
55 & person.com & 1574 & 0.20 & 47.8\\
56 & blogger.com & 1574 & 0.20 & 48.0\\
57 & gulli.com & 1568 & 0.20 & 48.2\\
58 & avast.com & 1551 & 0.20 & 48.4\\
59 & wikimedia.org & 1522 & 0.20 & 48.6\\
60 & amazonaws.com & 1497 & 0.19 & 48.8\\
61 & aol.com & 1472 & 0.19 & 49.0\\
62 & 194.79.31 & 1453 & 0.19 & 49.2\\
63 & 62.129.141 & 1444 & 0.19 & 49.4\\
64 & xtube.com & 1424 & 0.18 & 49.5\\
65 & movies.netbg.info & 1404 & 0.18 & 49.7\\
66 & oxedion.com & 1373 & 0.18 & 49.9\\
67 & flyupload.com & 1315 & 0.17 & 50.1\\
68 & veoh.com & 1265 & 0.16 & 50.2\\
69 & 61.183.15 & 1254 & 0.16 & 50.4\\
70 & mihd.net & 1226 & 0.16 & 50.6\\
71 & 21centurydns.com & 1225 & 0.16 & 50.7\\
72 & sekio.f28.us & 1214 & 0.16 & 50.9\\
73 & strategyinformer.com & 1201 & 0.16 & 51.0\\
74 & 194.246.114 & 1182 & 0.15 & 51.2\\
75 & cloak.ws & 1178 & 0.15 & 51.3\\
76 & google-analytics.com & 1168 & 0.15 & 51.5\\
77 & cupidbay.com & 1128 & 0.15 & 51.6\\
78 & xvideos.com & 1121 & 0.14 & 51.8\\
79 & 125.76.254 & 1120 & 0.14 & 51.9\\
80 & 64.246.54 & 1108 & 0.14 & 52.1\\
81 & bbc.co.uk & 1100 & 0.14 & 52.2\\
82 & putfile.com & 1069 & 0.14 & 52.3\\
83 & atdmt.com & 1067 & 0.14 & 52.5\\
84 & apple.com & 1057 & 0.14 & 52.6\\
85 & zshare.net & 1050 & 0.14 & 52.8\\
86 & rockyou.com & 1044 & 0.13 & 52.9\\
87 & mig33.com & 1027 & 0.13 & 53.0\\
88 & 4shared.com & 1024 & 0.13 & 53.2\\
89 & thunda.com & 1023 & 0.13 & 53.3\\
90 & rnbload.com & 1006 & 0.13 & 53.4\\
91 & mirror.ac.za & 1003 & 0.13 & 53.5\\
92 & com.com & 970 & 0.13 & 53.7\\
93 & gigasize.com & 948 & 0.12 & 53.8\\
94 & fadencreative.com & 947 & 0.12 & 53.9\\
95 & flixster.com & 938 & 0.12 & 54.0\\
96 & sendspace.com & 934 & 0.12 & 54.2\\
97 & news24.com & 933 & 0.12 & 54.3\\
98 & ifilm.com & 913 & 0.12 & 54.4\\
99 & filehippo.com & 913 & 0.12 & 54.5\\
100 & 38.98.61 & 904 & 0.12 & 54.6\\\hline
\end{tabular}


\end{multicols}
\caption{Popular organisational web sites by download. The download
  figures are given in Megabytes. 
\emph{Perc} gives the
  percentage of download from this organisation. \emph{Cum} gives the
cumulative percentage.}
\label{table:urlbymb}
\end{table}

What is obvious from this is the dominance of social networking sites
(the top 2 sites -- almost 30\% of usage are social networking sites),
followed by sports. Google commands about 3.7\% of all hits (this
would be across all Google services). News sites (at positions 22, 26,
33, 70, 83) score highly but collectively command less than 2\% of all
hits.  There are some sites for software download (Symantic and
Adobe). Wikipedia and Wikimedia are at positions 28 and 59. Two
university web sites also appear. Other than these latter two sites,
no obviously academic sites appear. JSTOR appears at position 140,
Springer at position 253. There are only two foreign universities in
the top 1000. There is some web browsing that is academic in nature,
and it may well be essential to the studies of the students involved,
but as a statistical phenomenon, web browsing should be viewed as a
social activity, just as participation in sports, student clubs or
parties.

Table \ref{table:urlbymb} shows the top 100 popular sites by download
size. Overall the picture is similar, though some obvious software
download sites (\url{adobe.com}, \url{macromedia.com},
\url{caltech.edu}) score highly. There may be some grey areas, and maybe some academics use Facebook
for teaching activities but it is obvious that the vast bulk of internet
download is used for social networking.

\subsection{Internet use by time of day}

\label{sec:use:time}
\begin{figure}[ht!]
\centering
\includegraphics{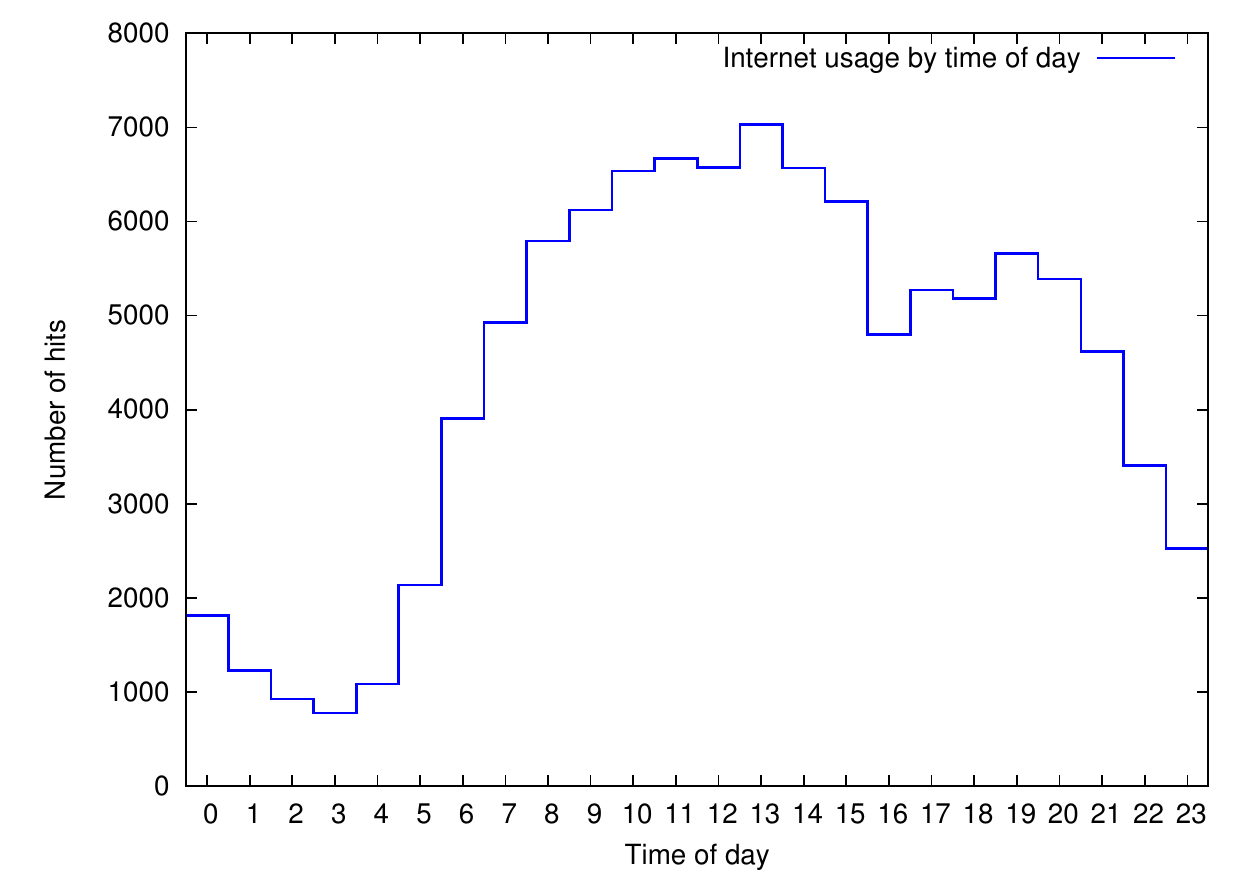}
\caption{Internet use by time of day}
\label{fig:usetime}
\end{figure}

Table~\ref{table:usetime} shows the use of internet by time of day.
This is shown graphically in Figure~\ref{fig:usetime}. As would be
expected, peak usage is around noon, but internet use remains high in
the early evenings.

\begin{table}[ht!]\centering
\begin{multicols}{3}
\begin{tabular}{|r|r|}\hline
Time & Number of hits\\\hline
0& 1809726 \\
1& 1230086 \\
2& 923822 \\
3& 774641 \\
4& 1085260 \\
5& 2135939 \\
6& 3906422 \\
7& 4923858 \\\hline
\end{tabular}

\columnbreak

\begin{tabular}{|r | r|}\hline
Time & Number of hits\\\hline
8& 5794114 \\
9& 6121647 \\
10& 6535347 \\
11& 6671627 \\
12& 6575327 \\
13& 7029927 \\
14& 6567500 \\
15& 6210050 \\\hline
\end{tabular}

\columnbreak

\begin{tabular}{|r|r|}\hline
Time & Number of hits\\\hline
16& 4797886 \\
17& 5272394 \\
18& 5182216 \\
19& 5662758 \\
20& 5388305 \\
21& 4615795 \\
22& 3409579 \\
23& 2526842 \\\hline
\end{tabular}
\end{multicols}
\caption{Internet use by time of day}
\label{table:usetime}
\end{table}

\subsection{Analysis of internet usage by sessions}

\label{sec:use:sessions}

Defining a session from proxy data is hard. A similar problem is
tackled in zoology research -- to define animal feeding sessions from
observational data. See \cite{owensmith82,sibly90} for a detailed
discussion of some approaches. We applied the ``broken stick'' model
of \cite{sibly90} which we now give a simplistic description.

When a foraging animal such as a Kudu eats, the gap between mouthfuls
could either be an intra-meal or inter-meal gap. Both gaps can be
modelled as Poisson processes: a fast one which describes gaps between
mouthfuls in a meal, and a slow one which describes gaps between the
last mouthful of one meal and the first of the next one. A simple
technique for determining this is to compute log-frequencies of gaps,
and then to plot the graph. In a pure one Poisson process model, the
graph should be linear. In a pure two Poisson process model, the fast
Poisson process will dominate the first part of the graph (small gaps)
and the slow Poisson process will dominate the last part of the graph,
and hence the graph should look like a ``broken stick'': a steeply
descending line at the left which at some point rapidly becomes a much
shallower line towards the right. The break-point is then a good
candidate for the gap between bouts. This break-point can either be
found by eye or using the appropriate technique -- see \cite{sibly90}
for more details.

\begin{figure}[ht!]
  \centering
  \includegraphics{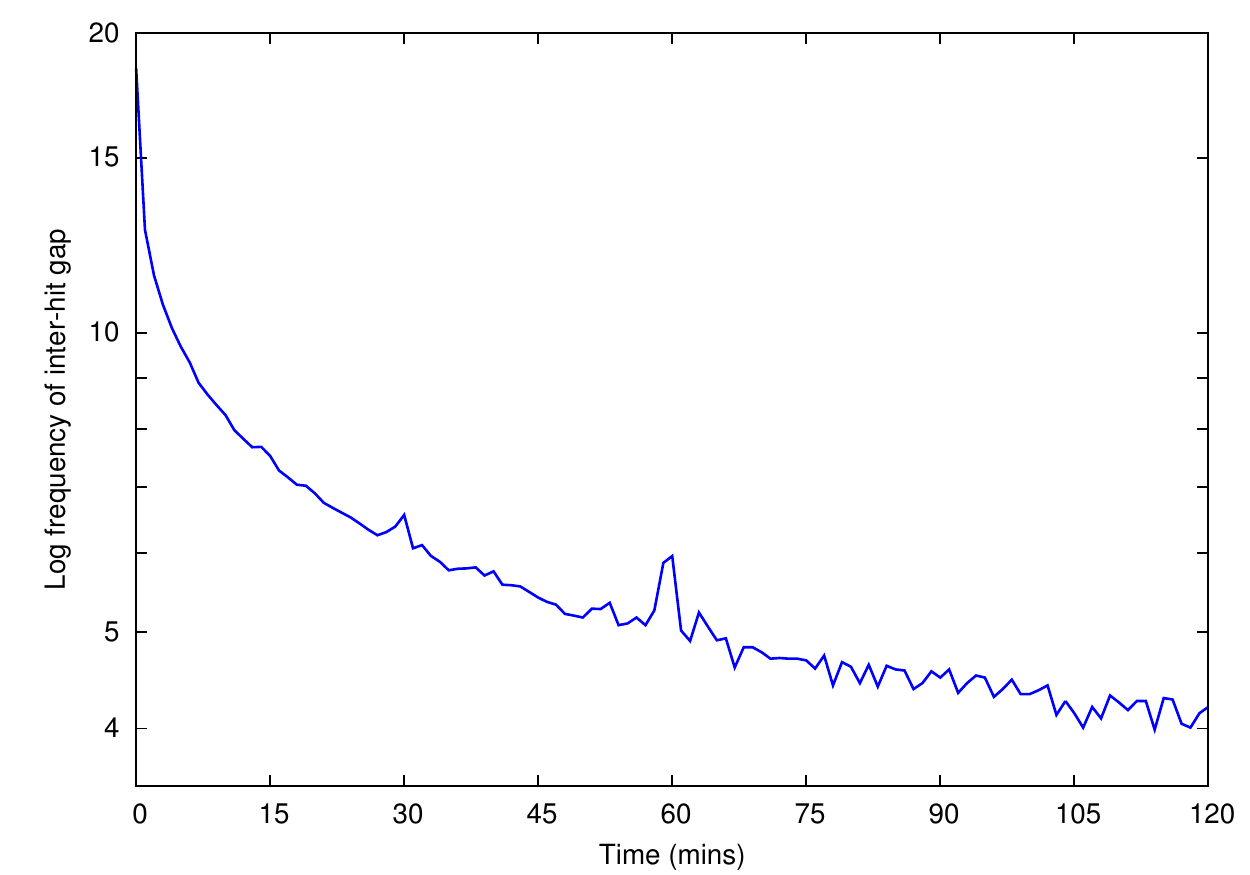}
  \caption{Plot of log-frequency distributions between user requests}
  \label{fig:sessionhisto}
\end{figure}

It is known that intra-session user behaviour is not Poisson, though
inter-session is  \cite{williamson2001}.  Nevertheless, we attempted
a similar analysis with our data to see whether an obvious cut-off
could be found. Figure \ref{fig:sessionhisto} shows the histogram of
gap lengths between individual users requests, binned into intervals
of minute lengths. It is clear from this analysis that there is no
obvious breakpoint. Experimenting with other distributions brought no
insight. 

This problem has been studied before and it is recognised as difficult
\cite{rugaber2006} and the best estimates are a gap of about 15
minutes is reasonable \cite{goker2000,murray2006}. We adopt the same,
and point out a serious methodological problem looking at the
logs. Many web pages (such as popular web-based email systems and
popular sports sites such as \url{cricinfo} and RSS feeds\footnote{We
  suspect that the high hit count for the BBC news site shown in our
  results is due to the fact that this site comes configured as a
  default RSS feed for Firefox. The peaks at the 30 minute and 60
  minute mark may also be due to auto-refresh.}) have
auto-refresh. Thus someone may not be using the internet at all, but
the logs would still record them as doing so.

\section{Academic Performance and Internet Usage Results}

\label{sec:results}

In this section, we compare academic performance to internet usage. We
primarily focus on internet usage as measured by number of bytes
downloaded. We also did the analysis based on a measurement of number
of hits and got broadly similar results --- for readability, the
details of the analysis based on number of hits can be found in the
appendix.

\subsection{Heavy versus light users}

Table \ref{table:acperfi} shows the performance of students in the
various categories (e.g. heavy vs light; all vs residence students). On the
whole, the table shows that the heavier users perform more poorly
than the lighter users both with respect to average mark and
performance index. As can be seen in Figure
\ref{fig:histo} the results are roughly bell shaped.

\begin{table}[ht]
  \centering
  \begin{tabular}{|l| r r|r r|} \hline
  Category & \multicolumn{2}{c|}{All students} & \multicolumn{2}{c|}{Residence students} \\\hline
  Very heavy& 49.6\% &$-4.21$  & 50.2\% &$-3.84$ \\
  Heavy     & 52.1\% &$-2.24$  & 51.6\% &$-2.90$  \\
  Light     & 54.6\% &$-0.54$  & 54.4\% &$-1.12$ \\
  Very light & 55.1\%&$+0.07$  & 54.6\% &$-1.53$ \\\hline
  \end{tabular}
  \caption{Academic Performance: Performance of students versus
    internet usage (bandwidth used). The figures in the table give the
  weighted average and performance index of the students in the
  different categories (e.g, heavy users in residence have an average of 51.6\%).}
  \label{table:acperfi}
\end{table}

We now do a detailed statistical comparison between light and heavy
users.  The descriptive statistics are shown below.

\begin{center}
\begin{tabular}{|l| r r|r r|}\hline
             &  \multicolumn{2}{c|}{Average mark} &
             \multicolumn{2}{c|}{PI}        \\\hline
All students &   Average  & Std Deviation & Average & Std Deviation\\\hline
Light        &   54.63\%  & 13.17         & $-0.535$& 12.48 \\
Heavy        &   52.10\%  & 11.43         & $-2.241$& 10.44 \\\hline
Res students &   Average  & Std Deviation & Average & Std Deviation\\\hline
Light        &   54.42\%  & 10.76         & $-1.123$& 10.41 \\
Heavy        &   51.64\%  & 10.74         & $-2.900$& 10.71 \\\hline
\end{tabular}
\end{center}

\paragraph*{All students}
A Welch two sample $t$-test to all the student data using a two tail
test\footnote{One tail is probably OK to use here since we did
  hypothesise a priori that heavier users would have poorer
  performance.} with different standard deviation shows that the
statistical difference based on average mark is highly significant
($p=0.0017$). The Wilcoxon rank sum test with continuity correction
yields $p<0.0001$. Based on performance index, this is significant on
the $t$-test ($p=0.024$) and Wilcoxon test gives $p=0.004$.

\paragraph*{Residence students}
A $t$-test yields a highly significant difference for the average mark
($p=0.02$) but not for the performance index ($p=0.13$).  The Wilcoxon
test results are ($p=0.011$) and ($p=0.167$).

\paragraph*{Summary}
Using the \emph{Wilcoxon rank sum test}, the difference between heavy
and light users is statistically significant with respect to average
mark for all students and students in residence. The difference in
performance index is statistically significant for all students but
not for residence students.  The difference between the statistical
significance results for average and performance index is due to the
latter smoothing out the difference in average marks between courses.

An analysis of academic performance versus number of hits yielded
similar results -- the details can be found in the appendix on
page~\pageref{sec:comp:acnumhits}.

We looked at residence students in more detail and examined
the difference between those students whose absolute use of the
internet was between 50 MB and 150 MB; and those who used between
500 MB and 5000 MB (roughly equal groups). The average mark (PI) of the
first group is 55.4\% ($-0.29$) for the second group
51.6\% ($-2.43$). The difference in average mark is highly statistically
significant ($p<0.01$), and the difference in performance index is
significant ($p<0.05$).

Thus, the results show that heavier users have poorer academic results
than lighter users. There is some statistical evidence to support
this. However, rather than using more statistical tests, we would
rather focus on the question of whether there is a practical
significant difference since 2\% does not seem to be a big difference,
especially in the face of confounding factors. We move to this
question now.

\subsection*{Distribution of marks}

The difference of 1\% may be trivial (13\% or 14\%; 67\% or 68\%) or
huge (49\% and 50\%) i.e., the difference between passing and failing
a course.  This section analyses the performance of users based on
the class of pass or fail.  First, we look at the pass/fail issue.

\begin{table}[ht]
  \centering
  \begin{tabular}{|l|r r|r r|r r|} \hline
    \multicolumn{7}{|c|}{All students in the study}\\
    \multicolumn{1}{|c}{}&
    \multicolumn{2}{c|}{All}&\multicolumn{2}{c|}{Very heavy
      users}&\multicolumn{2}{c|}{Very light users}\\\hline
    All results     & 9359  & 100\%   &   873 & 100\%  &  1096 & 100\%\\
    Number of passes& 7242  &  77.3\% &   258 & 70.6\% &   903 & 83.4\% \\
    Number of fails & 2117  &  22.6\% &   615 & 29.4\% &   193 & 17.6\%  \\\hline\hline
    \multicolumn{7}{|c|}{Students in residence}\\
    \multicolumn{1}{|c}{}   &
    \multicolumn{2}{c|}{All}&\multicolumn{2}{c|}{Very heavy
      users}&\multicolumn{2}{c|}{Very light users}\\\hline
    All results     & 3136  & 100\%   &   285 & 100\%  &   299 & 100\%\\
    Number of passes&  2386  &  75.9\% &   217 & 71.1\% &   211 & 76.2\% \\
    Number of fails &  757  &  24.1\% &   88 & 28.9\% &    66 & 23.8\%  \\\hline
  \end{tabular}
  \caption{Academic performance of very heavy and very light users --
    pass/fail outcomes in relation to bandwidth used across all courses (on average students register for 6.1 courses). }
  \label{table:acperfii}
\end{table}
Table \ref{table:acperfii} shows the association of internet usage and
passing or failing for very heavy and very light users. For each of
the students in the study we looked at the number of courses passed
and failed. We see that very heavy usage is associated with higher
failure rates. Interestingly, among residence students, very light
usage is associated with slightly higher failure rates than for light
usage. We conjecture this is due to very light usage being related to
the student effectively dropping out; for students not in residence,
very light usage does not necessarily indicate they have dropped out
since they may be using the internet at
home. Table~\ref{table:acperfiihit} in the appendix shows the same
thing by number of hits.

\begin{table}[t!]
  \centering
  \begin{tabular}{|l|r r|r r|r r|} \hline
   \multicolumn{7}{|c|}{All students in the study}\\
\multicolumn{1}{|c}{}& \multicolumn{2}{c|}{All}&\multicolumn{2}{c|}{Heavy users}&\multicolumn{2}{c|}{Light users}\\\hline
  All results     & 9359  & 100\%   &  2828 & 100\%  &  2816 & 100\%\\
  Number of passes& 7242  &  77.3\% &  2129 & 75.3\% &  2204 & 78.3\% \\
  Number of fails & 2204  &  22.6\% &   699 & 24.7\% &   612 & 21.7\%  \\\hline\hline
   \multicolumn{7}{|c|}{Students in residence}\\
\multicolumn{1}{|c}{}   & \multicolumn{2}{c|}{All}&\multicolumn{2}{c|}{Heavy users}&\multicolumn{2}{c|}{Light users}\\\hline
  All results     & 3136  & 100\%   &  1004 & 100\%  &   936 & 100\%\\
  Number of passes&  2386  &  75.9\% &   753 & 75.0\% &   728 & 77.8\% \\
  Number of fails & 757  &  24.1\% &   251 & 25.0\% &   208 & 22.2\%  \\\hline
  \end{tabular}
  \caption{Academic performance of heavy and light users -- 
    pass/fail outcomes in relation to bandwidth used across all courses.}
  \label{table:acperfiii}
\end{table}

More interestingly, Table~\ref{table:acperfiii} shows the performance for
heavy and light users. This also shows that heavier internet usage is
associated with higher failure rates. A $\chi^2$-test shows that this
is highly significant ($p<0.0001$ for all students, $p=0.034$ for residence
students).

\begin{figure}
  \centering
  \subfigure[All students in study]{\includegraphics[scale=0.62]{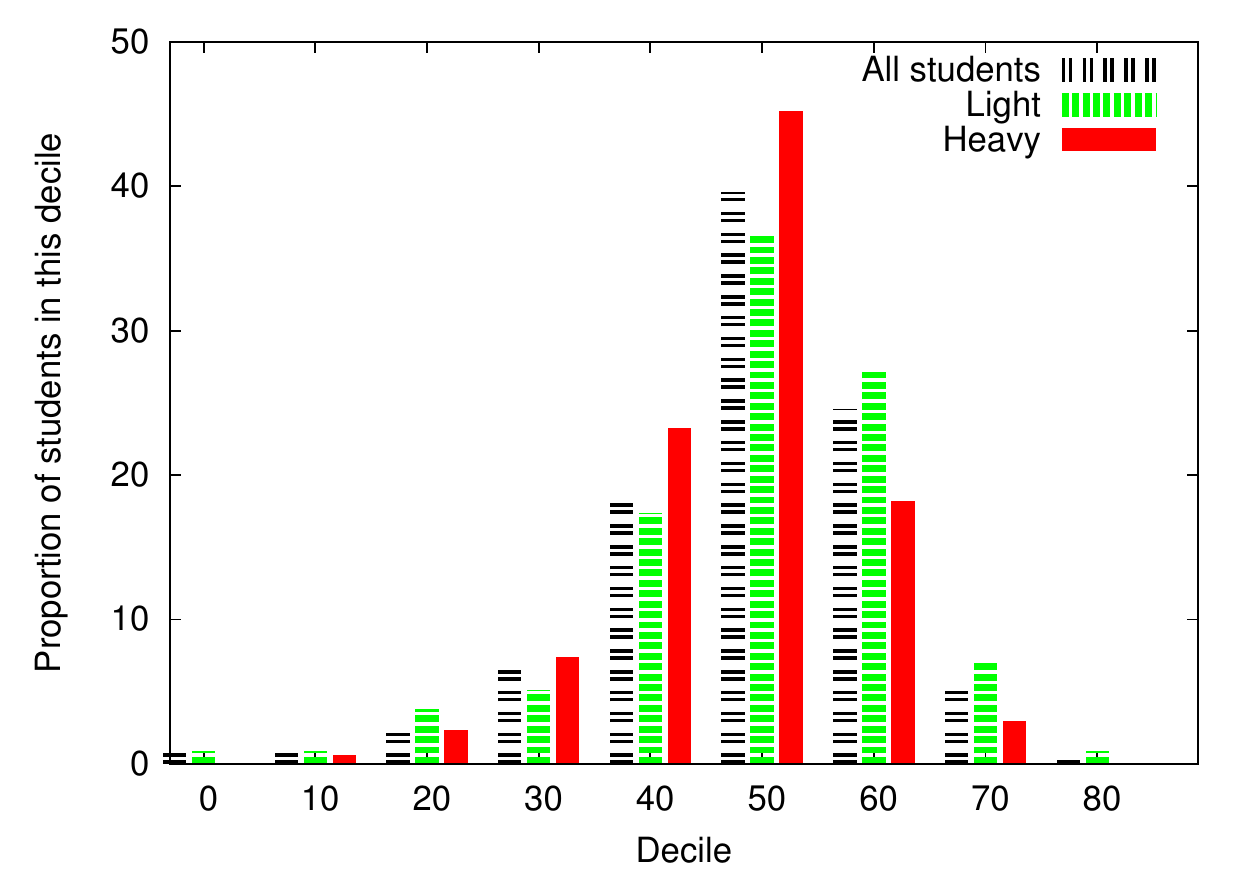}}
  \subfigure[Res students only]{\includegraphics[scale=0.62]{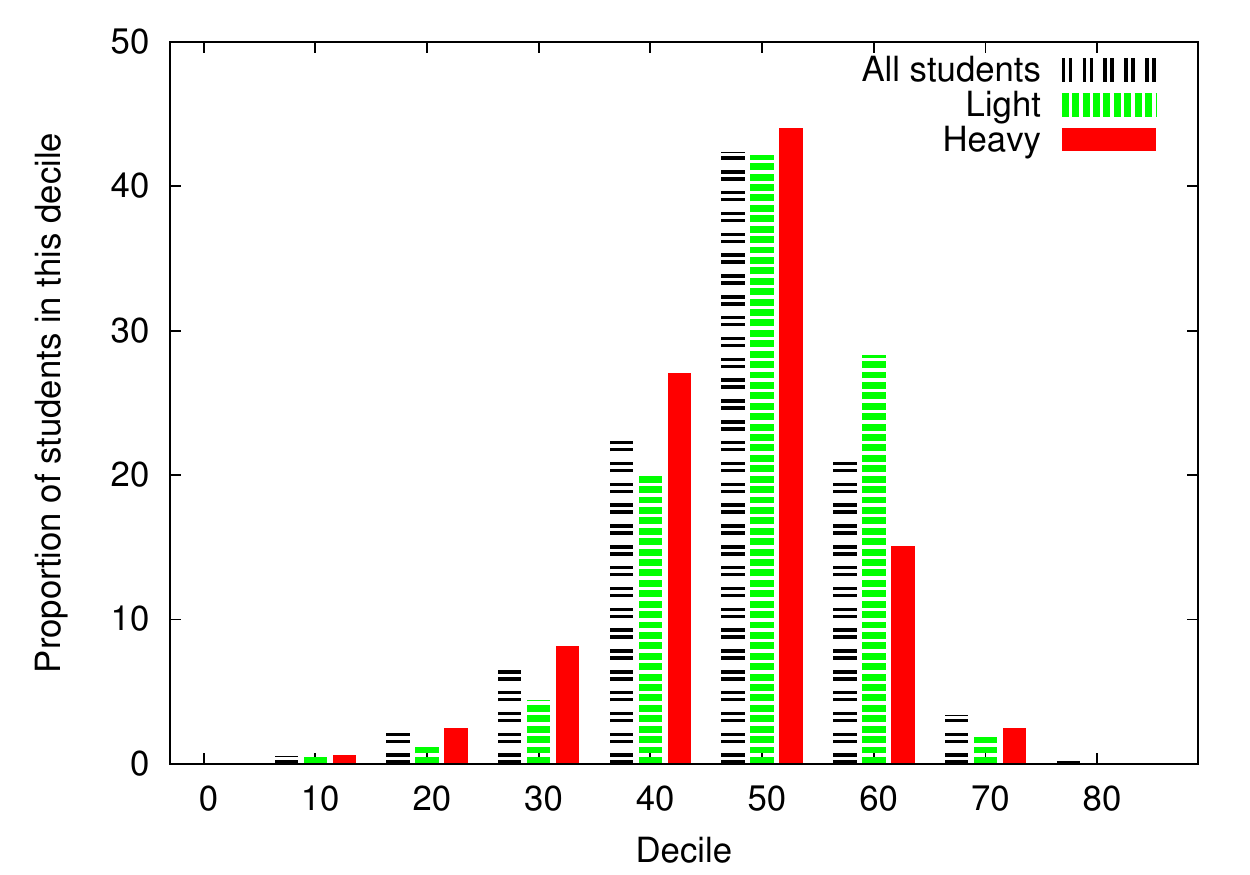}}
  \caption{Histogram of results -- (a) all students in the study, (b)
    residence students -- see Table \ref{table:resultshisto}}
  \label{fig:histo}
\end{figure}

Table \ref{table:resultshisto} and Figure \ref{fig:histo} take this
analysis further and show that not only is internet usage associated
with average pass/failure results of students but also between
different average categories of
pass. Table~\ref{table:resultshistcourses} shows the same result where
we look at results across all courses (not just averaging over students).
A similar result can be seen for the results by number of hits -- see
Table \ref{table:resultshistonet}.

\begin{table}[b!]\centering
\begin{tabular}{|l|r r r|r r r|}
\hline
           & \multicolumn{3}{c}{All students} & \multicolumn{3}{c|}{Res students}\\
Mark range & All & Light & Heavy & All & Light & Heavy \\
\hline
0-9 & 16    & 5   & 4   & 3   & 1  & 1 \\
10-19 & 13  & 4   & 3   & 3   & 1  & 1 \\
20-29 & 44  & 17  & 11  & 13  & 3  & 4 \\
30-39 & 105 & 24  & 34  & 35  & 7  & 12 \\
40-49 & 289 & 79  & 105 & 120 & 32 & 43 \\
50-59 & 610 & 169 & 210 & 226 & 67 & 70 \\
60-69 & 379 & 127 & 82  & 114 & 45 & 25 \\
70-79 & 79  & 33  & 14  & 18  & 3  & 4 \\
80-89 & 6   & 4   & 0   & 1   & 1  & 0 \\
\hline
\end{tabular}
\caption{Histogram of results based upon bandwidth usage. This histogram shows the distribution of average marks of students. A small difference in marks makes a larger difference in symbols.}
\label{table:resultshisto}
\end{table}

\begin{table}[htp!]\centering
\begin{tabular}{|l|r r r|r r r|}
\hline
           & \multicolumn{3}{c}{All students} & \multicolumn{3}{c|}{Res students}\\
Mark range & All & Light & Heavy & All & Light & Heavy \\
\hline
0-9 &    287  & 87  &  87 & 88   & 18 & 38 \\
10-19 &   59  & 26  &  15 & 19  &   6 &  5 \\
20-29 &  172  & 63  &  60 & 59   & 15 & 21 \\
30-39 &  686  & 190 & 218 & 259  & 76 & 80 \\
40-49 &  913  & 246 & 319 & 325  & 93 & 107 \\
50-59 & 3317  & 891 &1132 & 1211 & 347& 399 \\
60-69 & 2641  & 813 & 731 & 831  & 261& 270 \\
70-79 & 1000  & 366 &  41 & 280  & 96 & 71 \\
80-89 & 245   & 117 &   4 &  59  & 20 & 13 \\
90-100& 39    & 17  &   0 &   4  &  4 &  0 \\ 
\hline
\end{tabular}
\caption{Histogram of results based upon bandwidth usage. This histogram shows the distribution  of marks across courses. A small difference in marks makes a larger difference in symbols.}
\label{table:resultshistcourses}
\end{table}

The number of percentage points difference in the averages of the
groups of students is small (2-3 percentage points), but as
Tables~\ref{table:acperfii} (very heavy versus very light users) and
\ref{table:acperfiii} (heavy versus light users), show this reflects in
higher failure rates (a $\chi^2$ test shows that this is highly
statistically significant). Not only does this make a
difference in pass and failure rates, but also between classes of pass
-- fewer heavy users get good marks.

\subsection{Internet usage versus academic performance}

We are primarily interested in the effect of internet usage on
academic performance but it is interesting to explore the patterns of
usage of good and weak students.

First, we focus on the good students to see if their browsing habits
are significantly different to the others. If we define ``good'' as an
average of 65\%, we see that 50.2\% of good students are light or very
light internet users (where if internet usage were not a factor we
would expect 40\% of the good students to be lighter users), and
27.1\% of good students are heavy or very heavy users (a ratio of
light-to-heavy users of 1.86 -- by the way heavy and light users are
defined, we would expect the ratio to be 1:1). Applying the $\chi^2$
test, we get a $p$ value of $0.0026$.

A similar result can be found with other cut-offs for ``good''
students: at 60\% the ratio is 1.7:1, at 70\% it is 2.23:1, at the
75\% level 4.75 (at higher levels, the numbers do not allow statistical
tests to be used, but no very heavy user of the internet had an
average of more than 75\%).

Table \ref{table:goodpoort} gives another view of the results.  The
effect of a few massive downloads skews the results significantly and
the number of ``good'' students is relatively small. We see, however,
that over all students, good students (defined as those with an
average at least 60) download less than weak students.  On average
good students download 266 MB while those with average 40\% or less
have an average download of 972 MB. Even if we exclude the really big
downloaders who may skew the results we see the same pattern. If we
exclude students with more than 10000 MB download the difference is
266 to 645 MB (and 266 to 512 for a 5000MB cutoff).  Of course, as we
exclude more students we lose information too -- the cut-off of 5000
MB or 10000 MB is probably the fairest. For res students the situation
is similar, good students download 448MB on average compared to 1659MB
for weak students, and 448MB compared to 773MB for a 10000MB cutoff.

\begin{table}[ht]\centering
\begin{tabular}{|l| r r r r r|}\hline
                 & \multicolumn{5}{c|}{Cut-off} \\
Category                  &       None &10000&5000& 2000& 1000\\\hline
All students $\geq 60\%$: &       266  &266 &  266 & 200 &130\\
All students $\leq 40\%$: &       972  &645 & 512 & 287 &186\\\hline
Res students $\geq 60\%$: &       448  & 448&448 & 316  &197\\
Res students $\leq 40\%$: &      1659  & 773&687 & 456  &310\\\hline
\end{tabular}
\caption{The entries in the table show usage in MB. The cut-off
  columns shows what happens if we exclude students who download more
  than the given cut-off figure. It is important to use some filter
  since otherwise a few very big downloaders skew the overall
  results. However, even at a cut-off of 5000MB we see that good
  students download substantially less than weak students}
\label{table:goodpoort}
\end{table}

We performed an initial analysis of academic performance versus the
number of sessions and total session length. The results are
consistent with the analysis above -- for example the very heavy
users have an average mark of 50.4\%, the very light users an
average of 54\%. The average mark of users in the top 40\% of users
is 51.5\%, those in the bottom 40\% have an average of
54.4\%.  In view of the consistency  and the methodological problems
discussed above in terms of measuring sessions we felt that further
analysis would not shed more light.

\section{Other investigations}\label{sec:other}

We also conducted the following analyses, none of which showed
meaningfully different results from the general results.

\begin{enumerate}
\item \emph{Does time matter?} We had conjectured that students who did
  significant browsing at asocial hours (between 23:00 and 07:00) would
  be particularly prone to having poor academic results (either
  through being sleep deprived, or missing lectures to catch up
  sleep).  We did an initial test to see whether time of day made a
difference. We tabulated usage figures for those students who used the internet
at asocial times (AU) (23:00-07:00) and got the results shown in
Table \ref{table:asocial}. It is true that the people who were heavy asocial hours
  browsers had poorer results but the percentages were well in line
  with the results above.

\begin{table}[ht]
  \centering
  \begin{tabular}{|l l|}\hline
Top VH of AU &  average mark 49.0, PI 0.87 \\
Top H of AU & average mark 50.8, PI 0.90\\
Non-AU      & average mark 53.0, PI 0.93\\
Failure rate of VH/AU &  30.1\% (212/688)\\
Failure rate of AU    &  23.1\% (1539/6719)\\
Failure rate of non AU&  22.2\% (587/2640)\\\hline
   
  \end{tabular}
  \caption{Performance of students who work at asocial times as
    computed by number of hits}
  \label{table:asocial}
\end{table}

\item \emph{Does quality matter?} The difficulty of studying the effect of what is downloaded is that
there is such a wide variety of material that is downloaded. 
 Since the top two sites (almost
  30\% of hits) were social networking sites, we investigated whether
  heavier users of these sites were particularly prone to poor
  marks.  In doing this we picked the top two social
networking sites taken from Table \ref{table:urlbyhit}
(\url{facebook.com} and \studentvillage). Collectively we
call these SN users. Recall from Table~\ref{table:acperfii} that the failure rate across
all courses is 22.6\%. The failure rate for SN users (any usage at all)
is 22.9\%, the failure rate for the top 20\% of SN users was 25.3\% and
the failure rate for the top 10\% of SN users is 30\%.
Again, the difference between heavy and light users was in
  line with the results above. This is understandable since even if we
  exclude the top two sites, the vast majority of sites are still
  non-academic. There are many different ways to waste your time.

\item \emph{Does Discipline make a difference?} Do, for example, science students
    react differently to humanities students? Our initial study
  \cite{johnson2009} indicated that there might be differences (for
  example the average marks of Economics students with high usage was
  higher than those of lighter users). However, when we did this initial study
  we only had partial marks for the students. When we
  had the complete results, although the exact differences in average
  marks of heavy and light users differed, there was almost always a negative
  association, even if small. From the internet usage profile it is
  also obvious that the difference in internet usage cannot be
  explained by the academic requirements of the different courses,
  though availability of computers and psychological profile might
  have an impact.\\
\noindent \emph{Students registered for an second year economics course}
565 students. Heavy users have an average of 46.2\% and light users
have an average of 49.7\%.

\noindent \emph{Students registered for an second year maths course.} 
771 students. Heavy users have an average of 55.3\% and light users
have an average of 58.0\%.

\noindent \emph{Students registered for an second year maths course and
  not for a second year economics course}
681 students. Heavy users have an average of 56.7\% and light users
have an average of 58.9\%.

\noindent \emph{Students registered for an second year sociology,
  political science or anthropology and
  not for a second year economics course}
385 students. Heavy users have an average of 55.52\% and light users
have an average of 55.14\%. This was the one exception to the rule
that heavy users perform worse than light users

\noindent \emph{Students registered for a second year law course but
  not economics} 949 students. Heavy users have an average of 52.7\%
and light users have an average of 55.5\%.

\end{enumerate}

\section{Conclusion}

\label{sec:concl}

The results of our study show the following:
\begin{itemize}
\item The vast majority of internet browsing by University students is
  non-academic in nature. The web is primarily a social space for
  students. This does not mean that the web does not have an academic
  role, and in some areas it is transformative. However, if we want it
  to be transformative, it is unlikely to happen by accident.

\item Students with higher internet usage have a lower average than
  students with lower internet usage. The difference is small, but
  meaningful and statistically significant. This small difference in
  average marks is reflected in higher failure rates and fewer
  students with good marks.

\item The internet usage of good students has a significantly
  different profile to that of weak students. Light users are
  disproportionately represented among good students.
\end{itemize}

What lessons can be drawn, and what further research should be done?
We emphasise that with respect to the negative effect of internet
browsing, these are more symptoms of underlying problems rather than
primary causes.  

There is a natural inclination on seeing results like these to apply
measures to control access to certain web sites or classes of
material for one of two reasons: (1) To prevent students from wasting
their time and (2) To mitigate resource contention. While we have
some sympathy with the second reason, the first reason seems to miss
the point that there are underlying problems that should be tackled,
and that even though heavy internet usage is associated with poorer
marks, the difference is not that dramatic.

In our view the key issues are the following:

\begin{itemize}
\item An examination of the logs shows that a small group of students
  use the internet so much that it must be dysfunctional. This
  probably affects less than 5\% of students and can only be dealt
  with through pastoral care of students, which is often not possible
  in a large university. Detecting the problems is also
  hard. While it is possible to give real-time feedback to residence
  officials, for examples, on internet usage, this is problematic when
  it comes to issues such as privacy.

\item Time-planning and organisational skills. Improving students'
  abilities to organise themselves and improve their self-discipline
  is independently desirable. Our results show that the internet is
  another possible trap for students. This issue should explicitly be
  raised with students, and we believe that our results will be
  instructive. It is particularly instructive to see the difference is
  browsing patterns of good and weak students.
\end{itemize}

There were a number of relevant issues that our work only tangentially
dealt with. Future work should explore these. The ability to use the
internet \emph{effectively} is crucial. Often students (and staff) use
the internet and tools such as search engines and bibliographic
systems such as PubMed in the crude ways. There is a need to
improve these skills, though it is best for this to be done in an
integrated way. The use of the internet cannot be divorced from the
general ability to read, synthesise, understand and
write. Availability of the internet by itself does not help this -- it
may even have a negative effect. The question of e-learning, where
material and software systems are explicitly used to support teaching
is a separate issue.

\paragraph*{Ethics clearance:} This study was approved by the
\ifanonymous
the university ethics committee (protocol number suppressed).
\else
University of the Witwatersrand Human Subjects (Non-medical) Ethics
Committee, protocol number H080618.
\fi

\paragraph*{Thanks} 
\ifanonymous
We thank some people
\else.
We express our great thanks to Caroline
Fairbrother, head of the Wits Academic Information Systems Unit, for her
extensive help in providing and interpreting data. We owe her
big. We thank the Wits Computer \& Network Services (CNS) for their help
with proxy data, particularly Brett Geer, Umesh Bodalina and Hement Gopal.
\fi

\bibliography{refs}

\section*{Citation}

This is an extended version of a paper that appeared in the 2011
South African Computer Lecturer's Association Conference. If you reference this work please cite:
\begin{itemize}
\item S. Hazelhurst, Y. Johnson, I. Sanders. An empirical analysis of the relationship between web usage and academic performance in undergraduate students. \emph{Proceedings of the Annual Conference of the South African Computer Lecturer's Association}, Ballito, South Africa, July 2011, pp. 29-37.
\end{itemize}

\appendix

\section{Impact of number of hits made}

\subsection{Performance of students based on number of hits}

\label{sec:comp:acnumhits}

This section presents compares the academic performance of students
versus internet usage as measured by the number of hits.

\begin{table}[ht]
  \centering
  \begin{tabular}{|l|r r|r r|} \hline
\multicolumn{1}{|c|}{Category}   & \multicolumn{2}{c|}{All Students}&\multicolumn{2}{c|}{Residence Students}\\\hline
  Very heavy& 49.1\% &$-4.84$  &     49.5\% &$-4.53$  \\
  Heavy     & 51.8\% &$-2.43$  &     51.5\% &$-2.75$  \\
  Light     & 55.2\% &$-0.11$  &     54.4\% &$-1.48$ \\
  Very light & 55.5\%&$+0.99$  &     54.6\% &$+0.92$ \\\hline
  \end{tabular}
  \caption{Academic Performance Performance of students versus
    internet usage (number of hits). The figures in the table give the
  weighted average and performance index of the students.}
  \label{table:acperfihits}
\end{table}

Table \ref{table:acperfihits} -- based on hits rather than bandwidth used -- shows a similar trend to Table \ref{table:acperfi}.

\paragraph*{Statistical test:} All students average mark: Light users:
55.17/13.22. Heavy users 51.79/12.200. Using $t$-test and Wilcoxon highly
statistically significant ($p<0.0001$). Performance index: Light
users: $-0.114/12.7$; heavy users -2.425/11.2. Statistical tests
statistically significant ($p=0.003,0.00086$).

Residence students average mark: Light users: 54.40/11.20. Heavy users:
51.52/11.20. Using $t$-test and Wilcoxon significant ($p=0.022,
p=0.014$). Performance index: Light users -1.476/10.432. Heavy
users: $-2.750/11.12$. Not statistically significant ($p=0.29,0.49$).

\subsection{Pass/fail rates versus internet usage -- number of hits}

\begin{table}[ht!]
  \centering
  \begin{tabular}{|l|r r|r r|r r|} \hline
    \multicolumn{7}{|c|}{All students in the study}\\
    \multicolumn{1}{|c}{}&
    \multicolumn{2}{c|}{All}&\multicolumn{2}{c|}{Very heavy
      users}&\multicolumn{2}{c|}{Very light users}\\\hline
    All results     & 9359  & 100\%   &   918 & 100\%  &   971 & 100\%\\
    Number of passes& 7242  &  77.3\% &   657 & 70.9\% &   800 & 78.6\% \\
    Number of fails & 2117  &  22.6\% &   267 & 29.1\% &   171 & 21.4\%  \\\hline\hline
    \multicolumn{7}{|c|}{Students in residence}\\
    \multicolumn{1}{|c}{}   &
    \multicolumn{2}{c|}{All}&\multicolumn{2}{c|}{Very heavy
      users}&\multicolumn{2}{c|}{Very light users}\\\hline
    All results     & 3136  & 100\%   &   318 & 100\%  &   300 & 100\%\\
    Number of passes&  2386 &  75.9\% &   227 & 71.4\% &   229 & 76.3\% \\
    Number of fails & 757  &  24.1\% &    91 & 28.6\% &    71 & 23.7\%  \\\hline
  \end{tabular}
  \caption{Academic performance of very heavy and very light users --
    pass/fail outcomes in relation to number of hits. }
  \label{table:acperfiihit}
\end{table}

\begin{table}[htb]\centering
\begin{tabular}{|l|l l l|l l l|}
\hline
           & \multicolumn{3}{c}{All students} & \multicolumn{3}{c|}{Res students}\\
Mark range & All & Light & Heavy & All & Light & Heavy \\
\hline
0-9 & 16   & 6 & 6      & 2   & 1  & 1 \\
10-19 & 13 & 4 & 4      & 3   & 1  & 1 \\
20-29 & 44 & 14 & 13    & 13  & 2  & 4 \\
30-39 & 105 & 25 & 36   & 35  & 10 & 12 \\
40-49 & 289 & 71 & 99   & 120 & 36 & 41 \\
50-59 & 610 & 180 & 202 & 226 & 59 & 70 \\
60-69 & 379 & 131 & 84  & 114 & 42 & 21 \\
70-79 & 79 & 37 & 16    & 18  & 5  & 7 \\
80-89 & 6 & 5 & 1       & 1   & 1  & 0 \\
\hline
\end{tabular}
\caption{Histogram of results based upon number of hits. This histogram shows the distribution of marks. A small difference in marks makes a larger difference in symbols.}
\label{table:resultshistonet}
\end{table}

\end{document}